\documentclass[onecolumn,sort&compress,numbers]{els-mrw} % For numbered references Style 

\usepackage{amsmath,amssymb,amsfonts,amsthm,makeidx,graphicx}
\usepackage{txfonts}
\usepackage{helvet}
\usepackage{tikz}
\usetikzlibrary{arrows.meta,positioning,calc,patterns}
\pgfmathsetmacro{\dummy}{0}

\begin{document}

\chapter{Searching for New Physics with the Large Hadron Collider}\label{chap1}

\author[1]{Michael Spannowsky}

\address[1]{\orgname{Institute for Particle Physics Phenomenology}, \orgdiv{Durham University}, \orgaddress{Durham DH1 3LE, UK}}

\articletag{Chapter Article tagline: update of previous edition, reprint.}

\maketitle

\begin{abstract}[Abstract]
This chapter provides an introduction to collider phenomenology, explaining how theoretical concepts are translated into experimental analyses at the Large Hadron Collider (LHC). Beginning with the principles of collider operation and detector design, it outlines how collisions of protons are modelled through parton distribution functions, hard matrix elements, parton showers, and hadronisation. The discussion then turns to the reconstruction of physical objects and the definition of kinematic observables that expose the quantum numbers and dynamics of the underlying interactions. Special emphasis is placed on jet physics, including infrared- and collinear-safe algorithms, grooming and tagging techniques, and modern reconstruction approaches to jet substructure. The chapter introduces event selection strategies, object identification, and multivariate classification methods, before presenting the statistical framework underpinning modern collider analyses, from likelihood construction to hypothesis testing and uncertainty treatment. Three representative case studies, the Higgs discovery in the diphoton channel, high-mass dilepton resonance searches, and constraints on new physics through the Standard Model Effective Field Theory, demonstrate how these ingredients combine in end-to-end analyses. The chapter concludes with a perspective on future colliders and the growing role of open data and simplified likelihoods in enabling reinterpretation and global fits.
\end{abstract}

\begin{keywords}
 collider phenomenology\sep Large Hadron Collider\sep jets and substructure\sep effective field theory (SMEFT)\sep statistical interpretation and inference
\end{keywords}

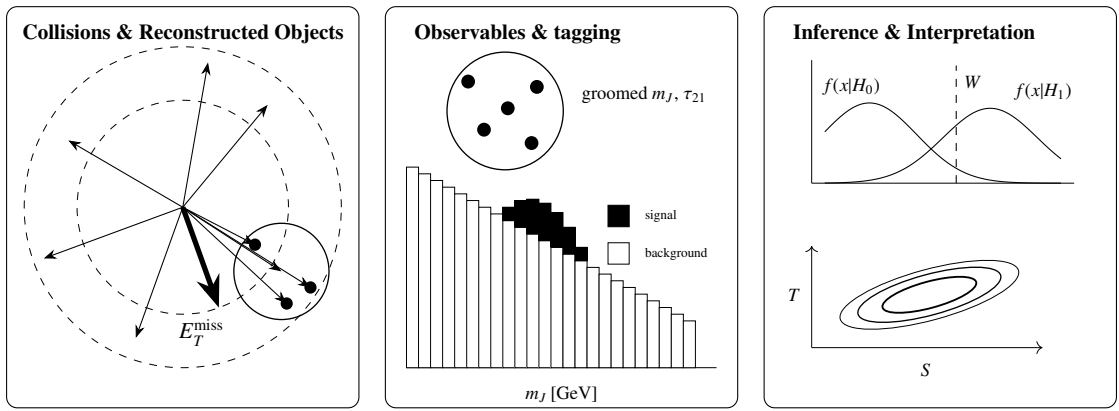
\begin{figure}[h]
  \centering
  \begin{tikzpicture}[x=1pt,y=1pt]
    \pgfmathsetmacro{\W}{420}    % total width
    \pgfmathsetmacro{\HH}{150}    % height
    \pgfmathsetmacro{\Pad}{8}
    \pgfmathsetmacro{\Sep}{10}
    \pgfmathsetmacro{\PW}{(\W-2*\Sep)/3}

    \draw[rounded corners=4pt, line width=0.4pt] (0,0) rectangle ({\PW},{\HH});
    \draw[rounded corners=4pt, line width=0.4pt] ({\PW+\Sep},0) rectangle ({2*\PW+\Sep},{\HH});
    \draw[rounded corners=4pt, line width=0.4pt] ({2*\PW+2*\Sep},0) rectangle ({3*\PW+2*\Sep},{\HH});

    % ===== Left panel: Collision & Objects =====
    \begin{scope}
      \clip (0,0) rectangle ({\PW},{\HH});
      
      \draw[->,>=Stealth] ({\PW*0.5},{\HH*0.5}) -- ++(80:55);
      \draw[->,>=Stealth] ({\PW*0.5},{\HH*0.5}) -- ++(150:50);
      \draw[->,>=Stealth] ({\PW*0.5},{\HH*0.5}) -- ++(200:56);
      \draw[->,>=Stealth] ({\PW*0.5},{\HH*0.5}) -- ++(250:52);
      \draw[->,>=Stealth] ({\PW*0.5},{\HH*0.5}) -- ++(50:50);
      % detector rings (stylised) 
      \draw[line width=0.3pt, dash pattern=on 3pt off 3pt] ({\PW*0.5},{\HH*0.5}) circle (40);
      \draw[line width=0.3pt, dash pattern=on 3pt off 3pt] ({\PW*0.5},{\HH*0.5}) circle (60);
      % large-R jet with three separated sub-blobs
      \draw[line width=0.5pt] ({\PW*0.78},{\HH*0.34}) circle (18);
      \fill ({\PW*0.78-10},{\HH*0.34+10}) circle (2.3);
      \fill ({\PW*0.78+11},{\HH*0.34-6}) circle (2.3);
      \fill ({\PW*0.78+2},{\HH*0.34-12}) circle (2.3);
      % arrow to jet as an object
      \draw[->,>=Stealth] ({\PW*0.5},{\HH*0.5}) -- ({\PW*0.78},{\HH*0.34});
      % arrows from center to each sub-blob
      \draw[->,>=Stealth] ({\PW*0.5},{\HH*0.5}) -- ({\PW*0.78-10},{\HH*0.34+10});
      \draw[->,>=Stealth] ({\PW*0.5},{\HH*0.5}) -- ({\PW*0.78+11},{\HH*0.34-6});
      \draw[->,>=Stealth] ({\PW*0.5},{\HH*0.5}) -- ({\PW*0.78+2},{\HH*0.34-12});
      % MET arrow
      \draw[->,>=Stealth,line width=2.0pt] ({\PW*0.5},{\HH*0.5}) -- ++(-70:40);
      % Precompute MET label position to avoid inline calc
      \pgfmathsetmacro{\metX}{\PW*0.35}
      \pgfmathsetmacro{\metY}{\HH*0.47}
      \pgfmathsetmacro{\metDX}{45*cos(-70)}
      \pgfmathsetmacro{\metDY}{45*sin(-70)}
      \node[anchor=west] at ({\metX+\metDX},{\metY+\metDY}) {$E_{T}^{\rm miss}$};
      % title
      \node[anchor=west,font=\small] at ({\Pad-5},{\HH-10}) {\textbf{Collisions \& Reconstructed Objects}};
    \end{scope}

    % ===== Middle panel: Observables & Tagging =====
    \begin{scope}[shift={({\PW+\Sep},0)}]
      \clip (0,0) rectangle ({\PW},{\HH});

      \draw[line width=0.5pt] ({\PW*0.34},{\HH*0.74}) circle (22);
      
      \fill ({\PW*0.34-14},{\HH*0.74+11}) circle (2.6);
      \fill ({\PW*0.34+12},{\HH*0.74+9}) circle (2.6);
      \fill ({\PW*0.34-8},{\HH*0.74-7}) circle (2.6);
      \fill ({\PW*0.34+10},{\HH*0.74-12}) circle (2.6);
      \fill ({\PW*0.34+1},{\HH*0.74+1}) circle (2.6);
      
      \node[font=\scriptsize,anchor=west] at ({\PW*0.34+26},{\HH*0.74+6}) {groomed $m_J$, $\tau_{21}$};

      \pgfmathsetmacro{\histX}{\Pad}
      \pgfmathsetmacro{\histY}{\HH*0.10}
      \pgfmathsetmacro{\histW}{\PW-2*\Pad}
      \pgfmathsetmacro{\histH}{66}
      \draw[line width=0.3pt] ({\histX},{\histY}) -- ({\histX+\histW},{\histY});
      \draw[line width=0.3pt] ({\histX},{\histY}) -- ({\histX},{\histY+\histH});

     \pgfmathsetmacro{\nbins}{24}
      \pgfmathsetmacro{\binw}{(\histW-8)/\nbins}

     \pgfmathtruncatemacro{\N}{\nbins}
      \foreach \i in {1,...,\N} {
        \pgfmathsetmacro{\xL}{\histX + (\i-1)*\binw}
        \pgfmathsetmacro{\xR}{\xL + \binw}
        \pgfmathsetmacro{\btop}{max(10, 75 - 2.5*(\i-1))}
        \pgfmathsetmacro{\yAxis}{\histY}
        \pgfmathsetmacro{\yTop}{\yAxis + \btop}
        \fill[white] ({\xL},{\yAxis}) rectangle ({\xR},{\yTop});
        \draw[line width=0.2pt] ({\xL},{\yAxis}) rectangle ({\xR},{\yTop});
      }

      \pgfmathsetmacro{\center}{ceil(\nbins/2)}
      \pgfmathtruncatemacro{\N}{\nbins}
      \foreach \i in {1,...,\N} {
        \pgfmathsetmacro{\xL}{\histX + (\i-1)*\binw}
        \pgfmathsetmacro{\xR}{\xL + \binw}
        \pgfmathsetmacro{\btop}{max(10, 75 - 2.5*(\i-1))}
        \pgfmathsetmacro{\yBase}{\histY + \btop}
        \pgfmathsetmacro{\sig}{max(0, 14 - (\i-\center)^2)}
        \pgfmathtruncatemacro{\sigok}{\sig > 0 ? 1 : 0}
        \ifnum\sigok>0
          \fill[black] ({\xL},{\yBase}) rectangle ({\xR},{\yBase + \sig});
          \draw[line width=0.2pt] ({\xL},{\yBase}) rectangle ({\xR},{\yBase + \sig});
        \fi
      }

      \pgfmathsetmacro{\legX}{\histX + 0.65*\histW}
      \pgfmathsetmacro{\legY}{\histY + 0.92*\histH}
      \pgfmathsetmacro{\ls}{8} 

      \draw[fill=black] ({\legX},{\legY}) rectangle ({\legX+\ls},{\legY-\ls});
      \node[anchor=west, font=\tiny] at ({\legX+\ls+3},{\legY-\ls/2}) {signal};

      \fill[white] ({\legX},{\legY-\ls-6}) rectangle ({\legX+\ls},{\legY-2*\ls-6});
      \draw[line width=0.2pt] ({\legX},{\legY-\ls-6}) rectangle ({\legX+\ls},{\legY-2*\ls-6});
      \node[anchor=west, font=\tiny] at ({\legX+\ls+3},{\legY-3*\ls/2-6}) {background};

      \node[font=\scriptsize] at ({\histX+\histW*0.5},{\histY-10}) {$m_J\,[\mathrm{GeV}]$};
      \node[font=\scriptsize, rotate=90] at ({\histX-14},{\histY+\histH*0.5}) {$\mathrm{d}\sigma/\mathrm{d}m_J$\,[arb.]};

      \node[anchor=west,font=\small] at ({\Pad},{\HH-10}) {\textbf{Observables \& tagging}};
    \end{scope}

    % ===== Right panel: Inference & Results =====
    \begin{scope}[shift={({2*\PW+2*\Sep},0)}]
      \clip (0,0) rectangle ({\PW},{\HH});
      % ===== Panel title =====
      \node[anchor=west,font=\small] at ({\Pad},{\HH-10}) {\textbf{Inference \& Interpretation}};

      \begin{scope}
        \pgfmathsetmacro{\topX}{\Pad + 10}
        \pgfmathsetmacro{\topY}{\HH*0.56}
        \pgfmathsetmacro{\topW}{\PW-2*\Pad-18}
        \pgfmathsetmacro{\topH}{46}
        \draw[line width=0.3pt] ({\topX},{\topY}) -- ++({\topW},0);
        \draw[line width=0.3pt] ({\topX},{\topY}) -- ++(0,{\topH});
        \pgfmathsetmacro{\muA}{\topX+0.22*\topW}
        \pgfmathsetmacro{\muB}{\topX+0.68*\topW}
        \pgfmathsetmacro{\sig}{0.18*\topW}
        \draw[domain=0:1,smooth,samples=40] 
          plot ({\topX+\topW*0.05+\topW*0.9*\x},{\topY+30*exp(-((\topX+\topW*0.05+\topW*0.9*\x)-\muA)^2/(2*\sig*\sig))});
        \draw[domain=0:1,smooth,samples=40] 
          plot ({\topX+\topW*0.05+\topW*0.9*\x},{\topY+28*exp(-((\topX+\topW*0.05+\topW*0.9*\x)-\muB)^2/(2*\sig*\sig))});
        \pgfmathsetmacro{\critX}{0.5*(\muA+\muB)}
        \draw[dashed] ({\critX+10},{\topY}) -- ({\critX+10},{\topY+\topH-2});
        \node[font=\scriptsize,anchor=south east] at ({\muA+7},{\topY+30}) {$f(x|H_0)$};
        \node[font=\scriptsize,anchor=south west] at ({\muB+6},{\topY+28}) {$f(x|H_1)$};
        \node[font=\scriptsize,anchor=west] at ({\critX+10},{\topY+\topH-8}) {$W$};
      \end{scope}

      \begin{scope}
        \pgfmathsetmacro{\botX}{\Pad + 10}
        \pgfmathsetmacro{\botY}{\HH*0.15}
        \pgfmathsetmacro{\botW}{\PW-2*\Pad-30}
        \pgfmathsetmacro{\botH}{38}
       
        \draw[line width=0.3pt,->] ({\botX},{\botY}) -- ++({\botW},0);
        \draw[line width=0.3pt,->] ({\botX},{\botY}) -- ++(0,{\botH});
        % Axis labels
        \node[font=\scriptsize,anchor=north] at ({\botX+0.5*\botW},{\botY-3}) {$S$};
        \node[font=\scriptsize,anchor=west] at ({\botX-12},{\botY+0.52*\botH}) {$T$};
        % Ellipse centers
        \pgfmathsetmacro{\ellX}{\botX+0.51*\botW}
        \pgfmathsetmacro{\ellY}{\botY+0.52*\botH}
        % 68% ellipse
        \draw[line width=0.7pt] ({\ellX},{\ellY}) ellipse [x radius={0.21*\botW}, y radius={0.13*\botH}, rotate=15];
        % 95% ellipse
        \draw[line width=0.5pt] ({\ellX},{\ellY}) ellipse [x radius={0.32*\botW}, y radius={0.20*\botH}, rotate=15];
        % 99% ellipse
        \draw[line width=0.3pt] ({\ellX},{\ellY}) ellipse [x radius={0.40*\botW}, y radius={0.25*\botH}, rotate=15];
      \end{scope}
    \end{scope}
\end{tikzpicture}
  \caption{Collider phenomenology at a glance. Left: proton--proton collisions produce final-state objects reconstructed as isolated photons and leptons, jets (including large-radius jets with internal substructure), and missing transverse momentum. Middle: observables and tagging methods, such as the groomed invariant jet mass, expose the quantum numbers and dynamics of the underlying interactions. Right: likelihood-based statistical inference turns measured spectra into discoveries, exclusions, and parameter constraints.}
  \label{fig:titlepage}
\end{figure}

\section*{Objectives}
\begin{itemize}
  \item Understand why colliders are uniquely powerful tools for discovering particles and testing the Standard Model.
  \item See how modern detectors turn collision debris into measurable objects that we can analyse.
  \item Learn how physicists design simple, robust observables to reveal the underlying quantum numbers and dynamics of a process.
  \item Appreciate what jets and event structure tell us about quarks, gluons, and heavy particles, without needing the technical details.
  \item Understand the idea of selecting events to enhance signals over backgrounds, and when more advanced classification can help.
  \item Grasp how data become results through likelihoods, uncertainties, significances, and limits.
  \item Recognise how shared tools and open data products make results reusable and comparable.
  \item See why future colliders and continual advances in theory and experiment are essential for deeper insights.
\end{itemize}

\section{Introduction}\label{intro}

Collider phenomenology is a central branch of particle physics that focuses on understanding, predicting, and interpreting the results of high-energy particle collisions. It serves as a bridge between the abstract world of theoretical models, such as the Standard Model and its extensions, and the concrete realm of experimental measurements performed at particle accelerators. The field’s primary aims are twofold: to rigorously test the Standard Model, which encapsulates our current best understanding of fundamental particles and their interactions, and to search for signatures of new phenomena that may point to physics beyond the Standard Model. These ambitious goals are motivated by some of the most profound open questions in science, including the nature of dark matter, the origin of the matter–antimatter asymmetry, the mechanism underlying electroweak symmetry breaking, and the possible unification of forces.

High-energy colliders provide a unique experimental environment to address these questions. By accelerating particles to velocities approaching the speed of light and colliding them at extreme energies, we probe interactions at distance scales far smaller than atomic nuclei, where quantum effects dominate and new particles or forces may reveal themselves. The outcome of these collisions is recorded as complex final states of known particles, from which we must reconstruct and infer the properties of the underlying fundamental interactions. This process demands a close interplay between theory and experiment, sophisticated detector technologies, and advanced statistical tools.

The Large Hadron Collider (LHC) at CERN, the most powerful collider ever built, sits at the forefront of experimental high-energy physics. By colliding protons at centre-of-mass energies up to 14 TeV and delivering unprecedented luminosities, the LHC enables exploration of both rare and high-mass processes. Its general-purpose detectors, ATLAS and CMS, are marvels of modern engineering, designed to reconstruct and measure a vast variety of final states: isolated leptons and photons, hadronic jets, and signatures of missing energy that may indicate the production of non-interacting particles. These capabilities make the LHC an exceptional tool for both precision tests of the Standard Model, such as detailed studies of the Higgs boson’s properties, and for the discovery of new physics scenarios, including supersymmetry, extra dimensions, and novel interactions.

Understanding how discoveries are made at colliders requires familiarity with several foundational concepts and methods, which this chapter introduces. We begin by reviewing the principles of particle colliders, contrasting them with fixed-target experiments and highlighting why colliders are uniquely suited to exploring the high-energy frontier. The concepts of centre-of-mass energy and luminosity are introduced, providing the kinematic and statistical context for interpreting collider data. We also discuss the design and function of modern detectors, whose layered structure enables the identification and measurement of diverse particles. The transformation from raw detector signals to reconstructed physics objects, such as isolated leptons, photons, jets, and missing transverse energy, is explained, laying the groundwork for the definition of robust, experimentally accessible observables.

Central to collider phenomenology is the use of kinematic observables that can efficiently separate signals of interest from Standard Model backgrounds. These observables are constructed from the four-momenta of reconstructed objects and are engineered to expose the quantum numbers and dynamics of the underlying interactions. For example, the invariant mass of particle pairs can reveal new resonances, while angular correlations and event shapes can distinguish between different production mechanisms or probe the spin and CP properties of new states. The design and selection of observables is intimately connected to the symmetries and structure of the underlying theory.

Jets, collimated sprays of hadrons arising from the fragmentation of quarks and gluons, are among the most ubiquitous and informative objects at hadron colliders. Their study has evolved into a rich subfield, encompassing not only the development of infrared- and collinear-safe clustering algorithms but also advanced techniques for grooming, tagging, and analysing jet substructure. These methods enable the identification of boosted heavy particles, such as top quarks, $W$, $Z$, and Higgs bosons, whose decay products may merge into single jets at high energies. Tagging algorithms further exploit displaced decay vertices and radiation patterns to distinguish heavy-flavour jets, hadronically decaying taus, and even to discriminate between quark- and gluon-initiated jets. Such techniques are crucial for both Standard Model measurements and searches for new physics.

A sophisticated theoretical framework underpins the interpretation of collider data. Because protons are composite objects, predictions for LHC observables require knowledge of the parton distribution functions, which encode the momentum structure of the proton in terms of quarks and gluons. The simulation of events proceeds through the calculation of hard matrix elements at parton level, followed by the modelling of QCD radiation in parton showers, and finally the non-perturbative process of hadronisation. Each stage introduces its own uncertainties and requires careful calibration against data. The resulting simulated events are then passed through detector simulations to enable direct comparison with experimental measurements.

Finally, the extraction of physics results from collider data relies on rigorous statistical inference. Likelihood-based methods are used to quantify the compatibility of data with different hypotheses, to set exclusion limits, and to measure parameters such as cross sections and coupling strengths. The treatment of systematic uncertainties through nuisance parameters, the construction of confidence intervals and hypothesis tests, and the use of control regions and auxiliary measurements are all essential elements of modern analyses. These statistical tools ensure that discoveries are robust and that results can be meaningfully compared across experiments and theoretical models.

Throughout this chapter, we aim to provide a self-contained introduction to these concepts, tailored to undergraduate and postgraduate students. By weaving together the principles of collider operation, detector design, theoretical modelling, kinematic analysis, jet physics, and statistical inference, we hope to equip readers with the conceptual framework necessary to understand how theoretical ideas are translated into experimental searches for new physics. Briefly, we will showcase concrete case studies, such as the discovery of the Higgs boson, resonance searches for new particles, and constraints on effective field theory operators, which will illustrate end-to-end analyses and highlight the interplay of ideas across the entire chain from theory to measurement.

\section{Collider Phenomenology Basics}\label{sec1}

\subsection{Historical Overview of Particle Colliders}\label{sec1:subsec1}

\begin{table}[t]
	\label{tab:colliders}
    \TBL{\caption{High-energy colliders of the past, present and future.}}
    {\begin{tabular*}{\textwidth}{@{\extracolsep{\fill}}@{}lllll@{}}
        \toprule
        \multicolumn{1}{@{}l}{\TCH{Collider}} &
        \multicolumn{1}{c}{\TCH{Site}} &
        \multicolumn{1}{c}{\TCH{Initial State}} &
        \multicolumn{1}{c}{\TCH{Energy}} &
        \multicolumn{1}{c}{\TCH{Discovery / Target}}\\
        \colrule
        SPEAR & SLAC & $e^+ e^-$ & 4 GeV & charm quark, tau lepton \\
        PETRA & DESY & $e^+ e^-$ & 38 GeV & gluon \\
        Sp\={p}S & CERN & $p \bar{p}$ & 600 GeV & W, Z bosons \\
        LEP & CERN & $e^+ e^-$ & 210 GeV & SM: electroweak and QCD tests \\
        SLC & SLAC & $e^+ e^-$ & 90 GeV & electroweak SM precision \\
        HERA & DESY & $e p$ & 320 GeV & quark/gluon structure of the proton \\
        Tevatron & FNAL & $p \bar{p}$ & 2 TeV & top quark and its properties \\
        BaBar / Belle & SLAC / KEK & $e^+ e^-$ & 10 GeV & quark mixing, CP violation \\
        LHC & CERN & $p p$ & 7/8/13/14 TeV & Higgs boson, electroweak sector, \\
		    &      &       &               & QCD, new physics \\
        \midrule
        \multicolumn{5}{c}{\textbf{Future Options}} \\
		\midrule
        ILC & --- & $e^+ e^-$ & 250--500 GeV (upgradable to 1 TeV) & precision electroweak and Higgs couplings \\
        CLIC & --- & $e^+ e^-$ & 380 GeV -- 3 TeV & precision electroweak and Higgs couplings \\
        FCC-ee / CEPC & --- & $e^+ e^-$ & 90--365 GeV & precision Higgs and electroweak physics \\
        Muon Collider & --- & $\mu^+ \mu^-$ & multi-TeV & discovery reach and electroweak physics \\
        FCC-hh & --- & $p p$ & 100 TeV & discovery of multi-TeV scale physics \\
        \botrule
    \end{tabular*}}{}
\end{table}

The record of achievements listed in Table~\ref{tab:colliders} shows the exceptional success of the global collider programme. Each generation of colliders has not only delivered headline discoveries but has also fundamentally reshaped our understanding of the building blocks of matter and the forces acting between them.  

Early colliders, such as SPEAR, were pivotal in discovering the charm quark and tau lepton~\cite{Perl:1975bf}, thereby establishing the existence of a new generation of matter particles. PETRA at DESY provided clear evidence for the gluon, confirming Quantum Chromodynamics as the theory of the strong interaction~\cite{TASSO:1979ckq}. The Sp\={p}S collider at CERN produced the W and Z bosons, delivering direct experimental confirmation of electroweak unification, one of the central achievements of the Standard Model~\cite{UA1:1983fsd}. LEP~\cite{ALEPH:2005ab,Rowson:1994zz}, operating as an extremely clean electron–positron collider, enabled unprecedented precision measurements of electroweak processes, while the SLC, with its polarised beams, offered unique complementary insights. HERA opened a window into the deep internal structure of the proton, probing quark and gluon distributions at small distance scales~\cite{H1:2015ubc}. The Tevatron pushed hadron collider techniques to new levels, culminating in the discovery of the top quark~\cite{Wagner:2010wd}. Finally, the LHC has taken collider physics to multi-TeV energies, discovering the Higgs boson~\cite{ATLAS:2012yve,CMS:2012qbp} and performing high-precision tests of Standard Model processes, while continuing to search for signs of new phenomena.  

Future colliders aim to build upon this remarkable legacy. The ILC~\cite{ILC-TDR-Physics-2013} and CLIC~\cite{CLIC-Summary-2018}, both linear electron–positron colliders, are designed for high-precision measurements of Higgs boson properties and electroweak interactions. Circular electron–positron colliders such as the FCC-ee~\cite{FCCee-CDR-EPJST-2019} and CEPC~\cite{CEPC-CDR-Vol2-2018} would produce unprecedented data samples, enabling measurements of exquisite precision across several energy thresholds. A muon collider~\cite{Delahaye:2019jvq}, representing a novel approach, could achieve multi-TeV collision energies in a relatively compact footprint because muons radiate far less energy in circular motion than electrons. Although muons are unstable, when produced and accelerated quickly, their lifetime is extended by relativistic time dilation, making them viable for high-energy collisions. At the highest energies, the proposed FCC-hh proton collider~\cite{FCChh-CDR-EPJST-2019} would provide discovery potential well beyond that of the LHC, offering sensitivity to entirely new mass scales and interactions. The particle physics community is actively engaged in long-term strategy processes to determine the optimal sequence and scope of these future facilities, balancing physics opportunities, technical feasibility and global resources.  

All of these machines rely on the elegant principle of accelerating charged particles using electric fields and steering them with magnetic fields. Protons and electrons have traditionally been preferred because they are stable and can be efficiently accelerated and stored, making them ideal for high-energy collisions. The exploration of muon colliders signals a readiness to innovate when compelling physics opportunities arise. This diversity of accelerator concepts demonstrates the adaptability and creativity of the field, which continues to evolve as it pushes toward deeper insights into the fundamental structure of nature.

\subsection{Particle Collider Principles}\label{sec1:subsec1}

Particle collider principles can be best understood by contrasting them with fixed-target or beam-dump experiments. 
In a beam-dump experiment, a single accelerated particle beam is directed onto a fixed target. 
Such an approach is experimentally simpler and allows dense target materials to produce high collision rates, 
which is advantageous for many intensity-frontier studies such as searches for weakly interacting particles or rare decays. 
However, in terms of the centre-of-mass energy available for producing new heavy particles, the kinematics are not optimal. 
For a projectile of energy $E$ and a stationary target of mass $m$, the total energy available for particle production 
in the centre-of-mass system scales as $M \sim \sqrt{2mE}$, 
meaning that even if one dramatically increases the projectile energy,  the produced mass threshold only grows with the square root of the beam energy. Because the centre of mass moves rapidly in the lab for a fixed‑target collision, most of the beam energy becomes kinetic energy of the centre‑of‑mass motion, not invariant mass available for producing new particles.

To see this explicitly, consider the production of a new state $M$ in the reaction $A + B \rightarrow M$. For a fixed-target (beam-dump) configuration, let the projectile of energy $E$ and mass $m_A$ move along the $z$-axis and collide with a stationary target of mass $m_B$.
The four-momenta before the collision are
\[
p_1 = (E, 0, 0, p), \qquad 
p_2 = (m_B, 0, 0, 0),
\]
with $p = \sqrt{E^2 - m_A^2}$.
The total squared centre-of-mass energy is
\[
s = (p_1 + p_2)^2 = m_A^2 + m_B^2 + 2 E m_B
\approx m_B^2 + 2 E m_B,
\]
where the last approximation assumes $E \gg m_A, m_B$. 
The maximum mass that can be produced in this configuration is
\[
M_\text{fixed target} = \sqrt{s} \approx \sqrt{2 m_B E},
\]
which grows only with the square root of the beam energy.

For a collider with two counter-propagating beams of equal energy $E$,
\[
p_1 = (E, 0, 0, p), \qquad 
p_2 = (E, 0, 0, -p),
\]
so that
\[
s = (p_1 + p_2)^2 = (E+E)^2 - (p - p)^2 = 4E^2.
\]
Hence, the maximum mass produced is
\[
M_\text{collider} = \sqrt{s} = 2E,
\]
which scales linearly with the beam energy and demonstrates the far greater efficiency of colliders
for producing heavy states compared to fixed-target experiments.

For this reason, high-energy particle physics has turned to colliders, in which two counter-propagating beams collide head-on. 
If both beams have equal energy $E$, the available energy for producing new states is 
$M \sim 2E$, which grows linearly with beam energy, making much more efficient use of the accelerator’s power. 
The cost of this configuration is that one can no longer use a dense stationary target; instead, two beams of finite 
particle density must be precisely focused and brought into collision, which imposes severe requirements on the luminosity 
of the machine.

The performance of a collider is quantified not only by its energy but also by its luminosity $\mathcal{L}$. 
Luminosity measures how many particles traverse the interaction region per unit time and per unit area and is given by
\[
\mathcal{L} = \frac{N_1 N_2 f}{A},
\]
where $N_i$ is the number of particles per bunch in each beam, 
$f$ is the frequency at which bunches cross, and $A$ is the transverse beam area at the interaction point. 
The number of events $N$ observed for a process with cross section $\sigma$ over a time period is 
\[
N = \sigma \int \mathcal{L} \, dt = \sigma \, \mathcal{L}_\text{int},
\]
where $\mathcal{L}_\text{int}$ is the integrated luminosity, typically reported in inverse femtobarns (fb$^{-1}$). 
For example, the LHC aims to deliver to each multi-purpose experiment, ATLAS and CMS, $\mathcal{L}_\text{int} \sim 3000~\mathrm{fb}^{-1}$ over its lifetime, 
sufficient to record billions of Higgs bosons and enable precision tests of the Standard Model.

There are further design choices between circular and linear colliders. 
Circular colliders continuously bend charged particles in a ring and allow repeated use of the same beam bunches, 
resulting in high collision rates. However, charged particles radiate energy when accelerated along curved paths, 
a phenomenon known as synchrotron radiation. 
For light particles such as electrons, synchrotron losses per revolution scale as $\Delta E \sim E^4 /(m^4 R)$, 
where $m$ is the mass of the accelerated particle and $R$ the bending radius. This shows explicitly that heavier particles, 
such as protons or muons, experience dramatically reduced energy losses compared to electrons at the same energy, 
which makes circular hadron colliders practical at high energies, while circular electron colliders become inefficient.
Linear colliders avoid synchrotron losses but use each bunch only once, so they face different luminosity challenges. 

Hadron colliders such as the LHC exploit the fact that protons are about 2000 times heavier than electrons and therefore 
suffer negligible synchrotron losses at TeV energies. The LHC accelerates counter-rotating bunches of protons in a 27 km circular tunnel and collides them at four interaction points equipped with large detectors such as ATLAS and CMS. This approach combines high centre-of-mass energy with high luminosity, enabling both the discovery of heavy new particles and precision studies of Standard Model processes. 
The use of bunches of protons also provides high collision frequencies while maintaining manageable radiation losses, a key reason why the current energy frontier relies on circular hadron colliders.

\subsection{Principles of Particle Detection}\label{sec1:subsec1}

Modern collider detectors are typically built as nearly hermetic cylindrical devices, with the particle interaction point located at the centre of the cylinder. This geometry provides coverage over almost the entire solid angle, allowing particles produced in high-energy proton–proton collisions to be detected and measured with high efficiency. Two prominent examples at the LHC are the ATLAS and CMS detectors, which share similar design principles but use different technical implementations.

Starting from the interaction point and moving outward, the first component is the \textit{inner tracking system}. It is designed to measure the trajectories of charged particles by recording their passage through multiple layers of silicon-based sensors placed in a magnetic field. From the curvature of these trajectories, one can infer the particle's charge and momentum. Surrounding the tracker is the \textit{electromagnetic calorimeter}, which measures the energy deposited by electrons and photons via the electromagnetic shower they produce when passing through dense material. Outside the electromagnetic calorimeter lies the \textit{hadronic calorimeter}, designed to absorb and measure the energy of hadrons such as protons, neutrons, and pions through hadronic cascades. The outermost subsystem is the \textit{muon system}, which identifies and measures the momentum of muons, the only charged particles typically able to traverse the calorimeters and exit the detector. A magnetic return yoke closes the field and provides support for the muon chambers. This layered structure enables the identification of particle types and the reconstruction of their kinematic properties with high precision.

The collisions at the LHC are produced by bringing into intersection two counter-rotating bunches of protons, each containing approximately $\mathcal{O}(10^{11})$ protons. When these bunches collide, a large number of particle interactions occur within a single bunch crossing, resulting in typically $\mathcal{O}(10^3)$ particles emerging from the interaction point. These particles create a large amount of radiation and a high particle multiplicity environment that the detectors must handle efficiently.

The data analysis does not rely directly on individual hits in the detector but on \textit{reconstructed physics objects}. At the LHC, these are defined as:
\begin{itemize}
  \item \textit{Isolated photons}, which deposit all their energy in the electromagnetic calorimeter without associated tracks in the tracker and are spatially separated from other hadronic activity.
  \item \textit{Isolated leptons} (electrons and muons), identified by their distinctive calorimetric and tracking signatures and separated from hadronic activity.
  \item \textit{Jets}, which are collimated sprays of hadrons originating from quarks and gluons, reconstructed by clustering algorithms applied to calorimeter energy deposits and tracks.
  \item \textit{Missing transverse energy} ($E_T^\text{miss}$), inferred from an imbalance of transverse momentum in the event and indicative of non-interacting particles such as neutrinos.
\end{itemize}

\begin{figure}[ht]
	\centering
	\includegraphics[width=.7\textwidth]{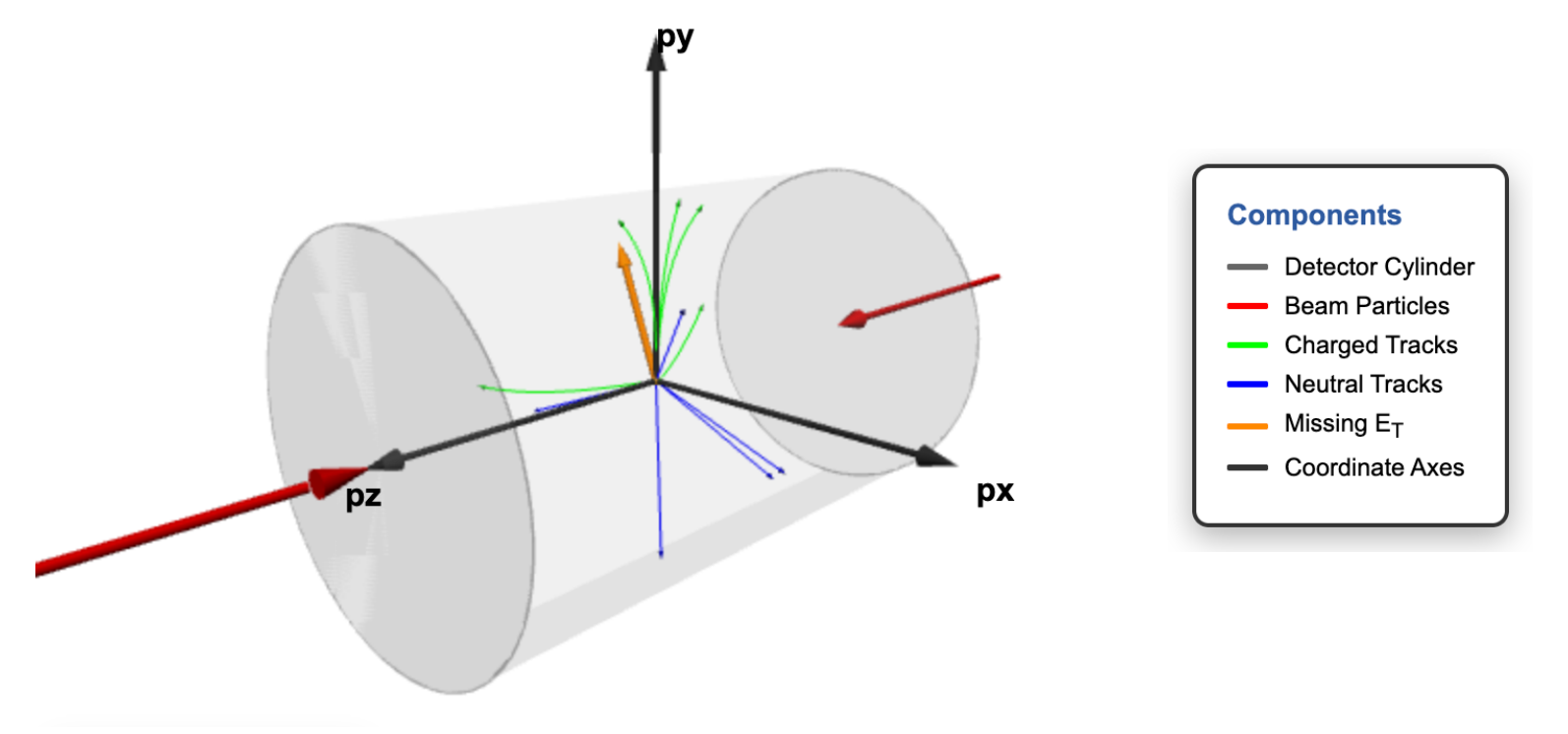}
	\caption{Schematic view of a collider detector. The green arrows correspond to charged particles, like electrons and muons, the blue arrows to neutral particles, like photons and neutrons. The orange arrow shows the calculated missing transverse energy of the event.}
	\label{fig:detector}
\end{figure}

The kinematics of these reconstructed objects are expressed in terms of coordinates adapted to the cylindrical symmetry of the detector. The azimuthal angle $\phi$ measures the rotation around the beam axis, and the rapidity $y$ or pseudorapidity $\eta$ parametrises the angle with respect to the beam direction:
\[
y = \frac{1}{2}\ln\left(\frac{E+p_z}{E-p_z}\right), 
\qquad
\eta = \frac{1}{2}\ln\left(\frac{|\vec p|+p_z}{|\vec p|-p_z}\right).
\]
The transverse momentum $p_T = \sqrt{p_x^2 + p_y^2}$ quantifies the momentum component perpendicular to the beam axis, which is the natural variable for hadron colliders because the initial state protons carry unknown longitudinal momentum fractions but have negligible transverse momentum.

A common way to visualise detector data is through \textit{lego plots}, which display the energy deposits in the calorimeter projected onto the $(\eta, \phi)$ plane, with the height representing transverse momentum $p_T$ or transverse energy $E_T$. These plots allow experimentalists to visually identify jets, photons, and other high-energy objects, providing a powerful tool for both online monitoring and offline physics analysis.

\begin{figure}[ht]
	\centering
	\includegraphics[width=.7\textwidth]{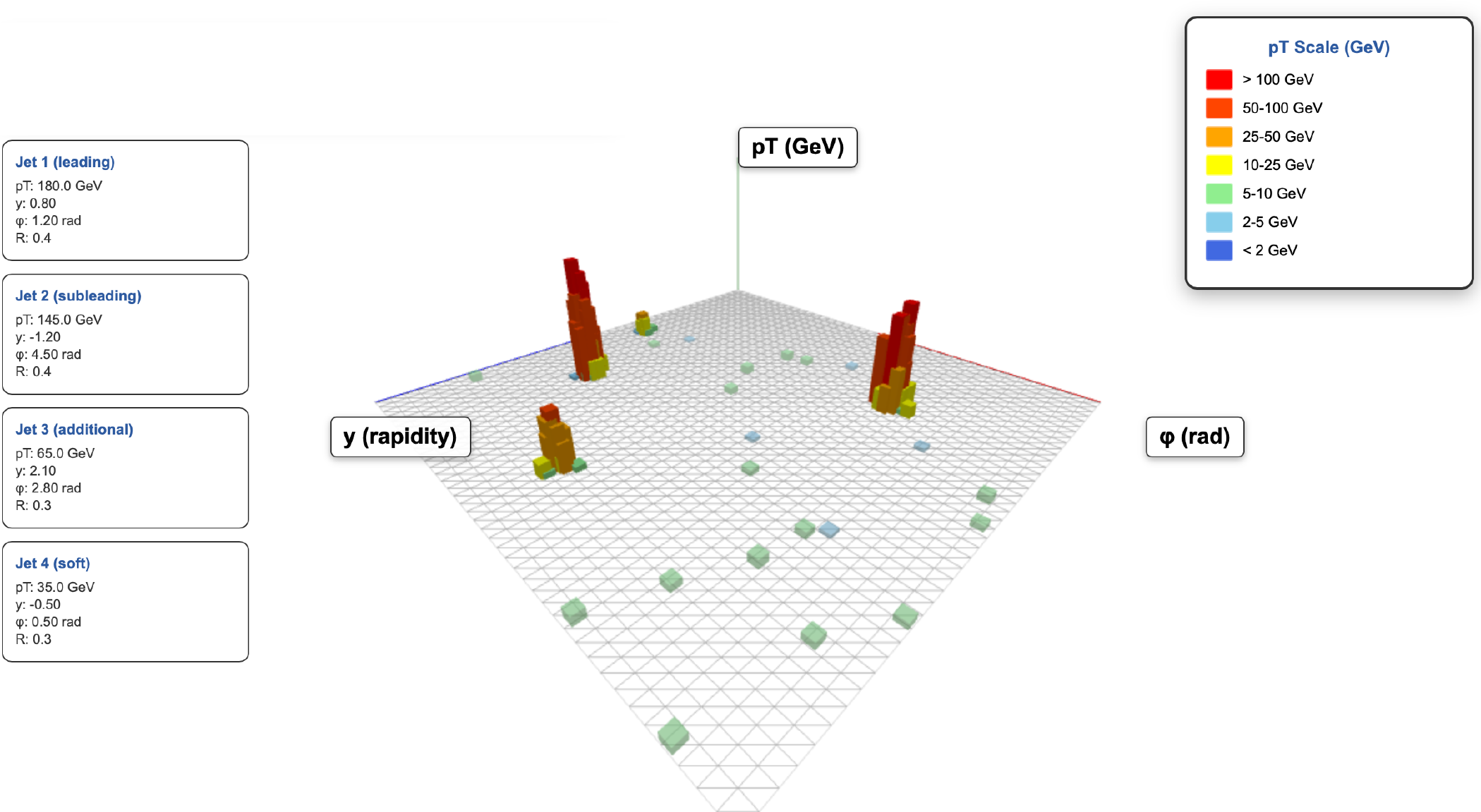}
	\caption{Lego plane plot of a multi-jet event at the LHC. The height of the towers represents the transverse momentum $p_T$, while the rapidity and azimuthal angle coordinates $(y, \phi)$ correspond to the particle directions.}
	\label{fig:lego}
\end{figure}

\subsection{The Theory Picture of Particle Collisions}\label{sec1:subsec1}

The interpretation of high-energy proton–proton collisions at the LHC is grounded in a robust theoretical framework that accounts for both the composite nature of the colliding particles and the multi-scale dynamics of Quantum Chromodynamics (QCD). Since protons are not elementary particles but rather bound states of quarks and gluons, any theoretical prediction for LHC observables must model both perturbative and non-perturbative aspects of QCD.

To achieve this, the simulation of LHC events is typically organised in a modular fashion. At the core of this framework lies the use of parton distribution functions (PDFs), which encode the probability of finding a parton (quark or gluon) carrying a given fraction \( x \) of the proton's momentum at a factorisation scale $\mu_F$. These functions are non-perturbative inputs extracted from global fits to experimental data and evolve according to the DGLAP equations. They allow us to write cross sections for hadron–hadron processes as convolutions of PDFs with partonic-level matrix elements.

Once the partons initiating the hard scattering process are selected via the PDFs, the next step is the calculation of the hard matrix element. This is the part of the simulation that is most directly tied to perturbative QCD and potentially to new physics. It involves computing the probability amplitude squared for a partonic subprocess such as $q\bar{q} \rightarrow Z \rightarrow \ell^+ \ell^-$ or $gg \rightarrow H$. These calculations are typically performed at leading order (LO), next-to-leading order (NLO), or even next-to-next-to-leading order (NNLO) in the strong coupling $\alpha_s$, depending on the precision required. The matrix elements determine the structure of the hardest interaction in the event. They are implemented either analytically or through automated tools such as \textsc{MadGraph}, \textsc{OpenLoops}, or \textsc{MCFM}.

After the hard interaction, one must account for the emission of additional QCD radiation from the incoming and outgoing partons. This is described by the parton shower, which simulates the sequential branching of quarks and gluons via processes like $q \rightarrow qg$, $g \rightarrow gg$, or $g \rightarrow q\bar{q}$. The parton shower resums the leading (and sometimes next-to-leading) logarithmic contributions to all orders in perturbation theory, capturing the collinear and soft enhancements of QCD. It produces a set of final-state partons that are still perturbative, though typically at low virtualities on the order of a few GeV.

Finally, these partons must be converted into observable hadrons. This is the role of hadronisation models, which describe the non-perturbative transition from coloured partons to colour-singlet hadrons. Several models exist, most notably the Lund string model and the cluster model, which are implemented in Monte Carlo event generators like \textsc{Pythia} and \textsc{Herwig}. The result is a fully exclusive final state composed of hadrons, which can then be passed to a detector simulation to produce observable signals.

Together, these four components, PDFs, hard matrix elements, parton showers, and hadronisation, form the theoretical backbone of LHC event simulation and are crucial for comparing theory predictions to experimental data.

In the following, we discuss each of these four components in more detail.

\begin{figure}[ht]
	\centering
	\includegraphics[width=.8\textwidth]{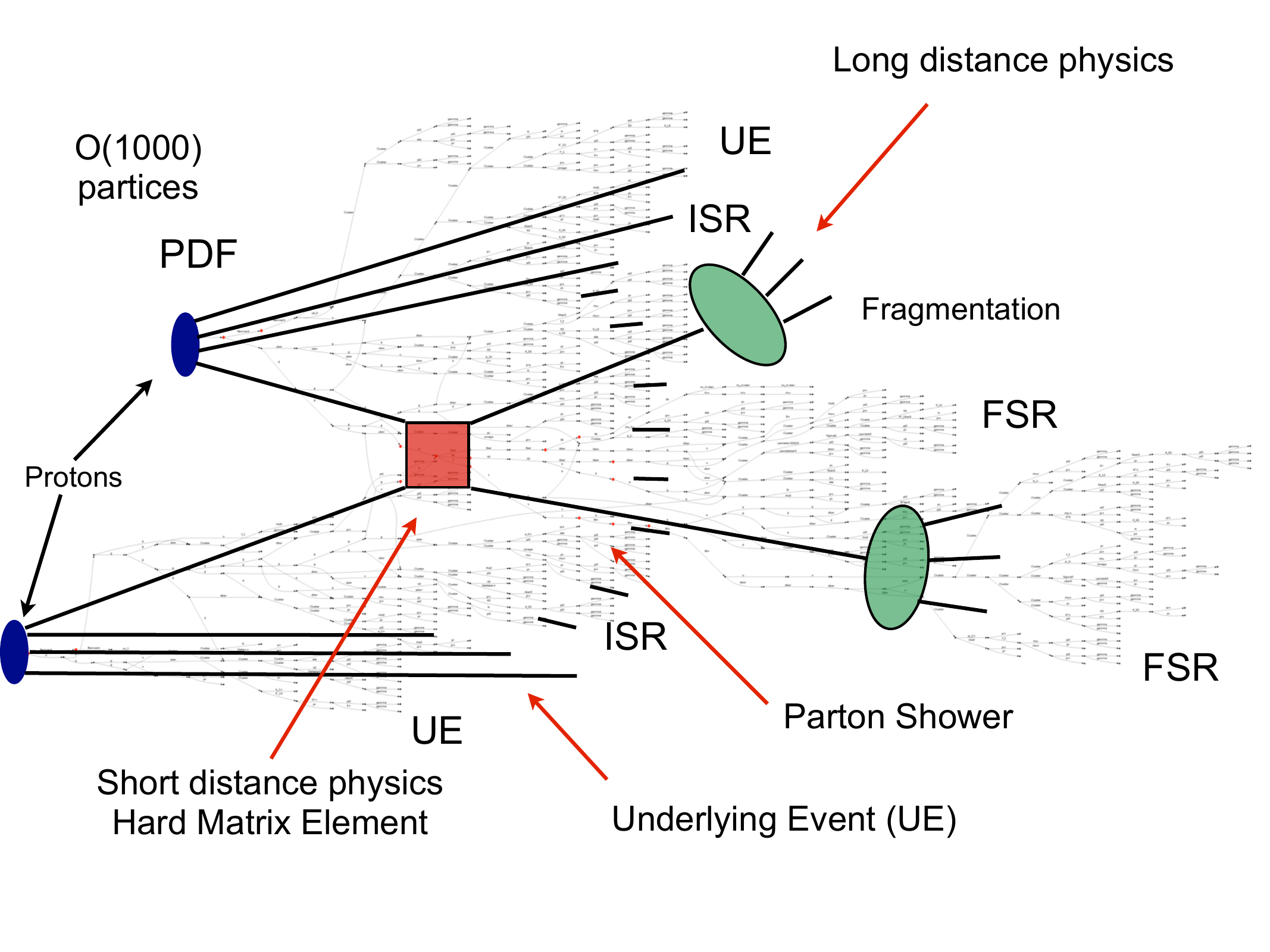}
	\caption{Theory picture overlaying a $pp \to t\bar{t}h$ collision event. The event is initiated by the hard scattering of two protons, represented by the blue blobs. The PDFs describe the momentum fractions $x_1$ and $x_2$ of the incoming partons. The hard matrix element $\hat{\sigma}_{ij \to X}$ is calculated for the partonic subprocess, which in this case is $gg \to t\bar{t}h$. The parton shower simulates the emission of additional quarks and gluons, leading to hadrons that subsequently decay into long-lived final objects that can be detected.}
	\label{fig:theorypicture}
\end{figure}

\subsubsection{Parton Distribution Functions}
\label{subsec:pdf}

The calculation of cross sections at hadron colliders requires understanding the momentum structure of the proton in terms of its constituents, quarks and gluons, collectively referred to as partons. PDFs provide the crucial link between the partonic-level theory and the hadronic-level observable. PDFs are non-perturbative objects that encode the probability density $ f_i(x,\mu_F)$ of finding a parton of type $i$ (quark, antiquark, or gluon) inside the proton, carrying a fraction $x$ of the proton’s longitudinal momentum when probed at the factorisation scale $\mu_F$.

In high-energy hadronic collisions, such as those at the LHC, the factorisation theorem of QCD allows us to write the cross section for a generic process as a convolution of PDFs with perturbatively calculable partonic cross sections:
\begin{align}
\sigma_{pp \to X} = \sum_{i,j} \int_0^1 dx_1 \int_0^1 dx_2\, f_i(x_1, \mu_F)\, f_j(x_2, \mu_F)\, \hat{\sigma}_{ij \to X}(x_1 x_2 s, \mu_F, \mu_R),
\end{align}
where $x_1$ and $x_2$ are the momentum fractions of the two partons from the incoming protons, $\mu_F$ is the factorisation scale, and $\mu_R$ is the renormalisation scale associated with the strong coupling $\alpha_s(\mu_R)$. The partonic cross section $\hat{\sigma}_{ij \to X} $ is computed using perturbative QCD.

The PDFs themselves cannot be calculated from first principles in perturbation theory due to their inherently non-perturbative nature. However, their scale dependence is governed by the DGLAP (Dokshitzer–Gribov–Lipatov–Altarelli–Parisi) evolution equations, which follow from the renormalisation group equations of QCD:
\begin{align}
\frac{d f_i(x, \mu^2)}{d \ln \mu^2} = \sum_j \int_x^1 \frac{dz}{z} P_{ij}\left(\frac{x}{z}, \alpha_s(\mu^2)\right) f_j(z, \mu^2),
\end{align}
where $P_{ij}(z)$ are the splitting functions that describe the probability of a parton $j$ emitting a parton $i$ and retaining a fraction $z$ of its momentum. These functions are calculable order by order in perturbation theory.

To obtain the initial conditions for the DGLAP evolution at a low factorisation scale \( \mu_0 \), PDFs must be extracted from experimental data via global fits. This is a highly non-trivial task that involves assuming a flexible functional form for the PDFs at the initial scale, parametrised by a set of free parameters. These parameters are then determined by fitting to a wide array of experimental measurements that are sensitive to different combinations of parton flavours and momentum fractions. Among the most important sources of data are deep inelastic scattering (DIS) measurements from HERA, which provide clean access to the quark and gluon content of the proton at varying values of Bjorken $x$ and $Q^2$ (see e.g. H1 and ZEUS combined results in \cite{H1:2009pze}). Drell–Yan production processes in proton–proton and proton–antiproton collisions, such as those measured at the Tevatron and LHC, offer complementary information on valence and sea quark distributions \cite{Moreno:1990sf, CMS:2013pzl}. Inclusive jet production at hadron colliders probes the gluon distribution at high scales and moderate to large $x$, thanks to the large momentum transfers involved \cite{D0:2000dzr, ATLAS:2011juz}. Precision measurements of electroweak boson production and top-quark pair production at the LHC further constrain the flavour decomposition and high-scale evolution of PDFs \cite{ATLAS:2011qdp, CMS:2019nrx}. The combination of these diverse datasets ensures a robust and comprehensive determination of the proton's partonic content across a wide kinematic range.

This procedure is carried out by several collaborations, including NNPDF, CT, and MSHT, each using different strategies and datasets \cite{Ball:2022qks, Hou:2019efy, Bailey:2020ooq}. The NNPDF collaboration, for instance, uses neural networks as unbiased function approximators and employs Monte Carlo methods to estimate PDF uncertainties. These uncertainties play a vital role in estimating the precision of theoretical predictions.

The output of such a PDF fit is a set of functions $f_i(x, \mu)$ for all relevant partons and for a range of scales $\mu$, together with error estimates. These PDFs are then made publicly available through libraries such as LHAPDF \cite{Buckley:2014ana}, which interface seamlessly with Monte Carlo generators and analytic calculation tools. Prominent general-purpose event generators that make use of these PDFs include \textsc{Pythia~8} \cite{Sjostrand:2014zea}, \textsc{Herwig} \cite{Bellm:2015jjp}, and \textsc{Sherpa} \cite{Gleisberg:2008ta}.

Hence, PDFs are fundamental objects in the theoretical prediction of collider observables. They encode our current knowledge of the proton’s internal structure and are indispensable for connecting high-energy QCD calculations to experimental measurements.

\subsubsection{Hard Matrix Element Calculation}
\label{subsec:mem}
At the core of theoretical predictions for processes at the LHC lies the calculation of hard scattering matrix elements. These represent the squared amplitudes for partonic subprocesses, such as $q\bar{q} \to Z$, $gg \to H$, or more complicated multiparton processes like $pp \to t\bar{t} + \text{jets}$. Such matrix elements are computed within the framework of perturbative QCD, using Feynman diagrams and the standard rules of quantum field theory.

The perturbative expansion of the cross section is organised in powers of the strong coupling constant $\alpha_s$. Schematically, it reads
\[
\sigma = \sigma_{\text{LO}} + \alpha_s ~\sigma_{\text{NLO}} + \alpha_s^2 ~\sigma_{\text{NNLO}} + \cdots
\]
where the leading-order (LO) term corresponds to the squared tree-level amplitude. The next-to-leading order (NLO) contribution includes one-loop virtual corrections and real emission of an extra parton. The next-to-next-to-leading order (NNLO) term adds two-loop virtual corrections and double-real emissions. These higher-order contributions are essential for precision phenomenology and for reducing theoretical uncertainties.

Calculating the matrix elements beyond LO is technically challenging, largely due to the appearance of divergences. Ultraviolet (UV) divergences arise from short-distance behaviour in loop integrals and are removed by renormalisation, introducing the renormalisation scale $\mu_R$. Infrared (IR) divergences arise from soft or collinear emissions and are regulated dimensionally. These divergences cancel between virtual and real-emission contributions when computing infrared-safe observables. However, to perform the cancellation explicitly in numerical computations, sophisticated subtraction schemes are required.

Subtraction methods systematically isolate and subtract the divergent parts of the phase space integrals. Examples include the Catani-Seymour dipole subtraction method, antenna subtraction, and the $q_T$-subtraction formalism~\cite{Catani:2007vq, Campbell:2006xx, Catani:2009sm}. Numerical implementations of these techniques are embedded in tools such as \textsc{MCFM}~\cite{Campbell:2010ff}, \textsc{NNLOJET}~\cite{Currie:2014pka}, and others. An alternative approach, based on sector decomposition, decomposes the integrals into singular and regular parts before performing the numerical integration. This method is implemented in the \textsc{SecDec} program and has proven valuable for NNLO and even N$^3$LO calculations in both QCD and electroweak theory~\cite{Borowka:2015mxa, Borowka:2017idc}.

Advances in recent years have been achieved through the use of unitarity-based and on-shell methods for loop amplitudes. Instead of computing all Feynman diagrams individually, these methods reconstruct loop amplitudes from their analytic structure. This dramatically reduces computational complexity. These techniques are particularly well-suited for automation and have been implemented in tools such as \textsc{OpenLoops}~\cite{Cascioli:2011va}, \textsc{GoSam}~\cite{Cullen:2011ac}, and \textsc{MadLoop}~\cite{Hirschi:2011pa,Bern:2007dw}.

Automation of hard matrix element generation, both at tree level and at one-loop level, has revolutionised the field. Tools like \textsc{MadGraph5\_aMC@NLO}~\cite{Alwall:2014hca}, \textsc{Sherpa}, and \textsc{Herwig} include interfaces for matrix element generation and allow users to simulate complex multi-particle final states with NLO accuracy in a largely automated manner.

At NNLO, the field remains highly technical. Several landmark calculations such as $pp \to H$~\cite{Anastasiou:2005qj}, $pp~\to Z$~\cite{Melnikov:2006di}, and $pp \to t\bar{t}$~\cite{Czakon:2015owf} have now reached fully differential NNLO precision. A numerical subtraction method is implemented in the \textsc{SecDec} program and has proven valuable for NNLO and even N$^3$LO calculations in both QCD and electroweak theory~\cite{Heinrich:2021dbf}.

The calculation of hard matrix elements at the LHC requires a combination of analytic insight and numerical techniques. These computations form the foundation of modern collider phenomenology and must be performed with high accuracy to match the experimental precision achieved at the LHC.

\subsubsection{Parton Shower Evolution}
\label{subsubsec:partonshower}

The parton shower stage connects the hard scattering process (see Sec.~\ref{subsec:mem}) with the non-perturbative hadronisation step (Sec.~\ref{subsubsection:hadronisation}). It also complements the structure encoded in the PDFs, introduced in Sec.~\ref{subsec:pdf}, by simulating additional perturbative QCD radiation in the final state. While the hard matrix element describes the short-distance partonic interaction at a fixed order in perturbation theory, the parton shower resums the dominant logarithmic contributions from soft and collinear emissions to all orders.

The parton shower is based on the universal infrared structure of QCD amplitudes. In the limit where a parton splits into two collinear partons or emits a soft gluon, the matrix element factorises into a reduced matrix element and a universal splitting function. These limits form the basis of a probabilistic cascade, whereby partons undergo successive branchings described by $1 \to 2$ splittings of the type $q \to qg $, $g \to gg $, or $g \to q\bar{q}$. The probability for a splitting is given by the DGLAP splitting functions $P_{a \to bc}(z)$, where $z$ denotes the energy fraction carried by one of the daughters.

The evolution variable of the shower can vary between implementations. Popular choices include the virtuality $Q^2$, the transverse momentum $p_T$, or the angle $\theta$ of the splitting. The emission probability between two scales $Q^2_{\text{max}}$ and $Q^2_{\text{min}}$ is governed by the Sudakov form factor:
\[
\Delta(Q^2_{\text{max}}, Q^2_{\text{min}}) = \exp\left( - \int_{Q^2_{\text{min}}}^{Q^2_{\text{max}}} \frac{\mathrm{d}q^2}{q^2} \int \mathrm{d}z \, \frac{\alpha_s}{2\pi} P(z) \right),
\]
which represents the probability for no resolvable emission in that range.

The parton shower resums leading-logarithmic (LL) terms of the form \( \alpha_s^n \log^{2n}(Q^2/q_0^2) \), where \( q_0 \) is an infrared cutoff. In more advanced formulations, such as dipole or antenna showers~\cite{Giele:2007di, Schumann:2007mg}, next-to-leading-logarithmic (NLL) accuracy is achieved by explicitly including angular correlations and spin information. Dipole showers model the radiation pattern as emitted from colour dipoles rather than individual partons, which ensures better matching to the soft and collinear structure of QCD.

Furthermore, the parton shower must be matched to the hard matrix element in such a way as to avoid double-counting emissions already included in the fixed-order calculation. This matching is done using techniques such as the MC@NLO or POWHEG formalism~\cite{Frixione:2002ik, Nason:2004rx}, which consistently combines NLO matrix elements with parton showers while preserving total rates and leading-logarithmic accuracy.

Most modern event generators implement both initial-state and final-state showers. Initial-state radiation (ISR) evolves backwards from the hard process toward the incoming hadrons and incorporates the effects of the PDFs. Final-state radiation (FSR) evolves forward from the outgoing hard partons. Colour coherence effects and recoil strategies are also implemented to preserve momentum conservation and the correct radiation pattern.

In addition to matching fixed-order calculations to parton showers, it is often necessary to merge matrix elements of varying final-state multiplicities with the shower evolution. This is particularly important for processes with many hard jets. The CKKW merging algorithm~\cite{Catani:2001cc} and its variants, such as CKKW-L~\cite{Lonnblad:2001iq} and MLM~\cite{Mangano:2006rw}, provide prescriptions for combining multiple LO or NLO matrix elements with parton showers without double counting. These algorithms introduce a merging scale to separate hard matrix element emissions from softer shower emissions, and employ Sudakov reweighting to ensure a smooth transition between the two regimes.

Parton shower algorithms are implemented in codes such as \textsc{Pythia~8}, \textsc{Herwig~7}, and \textsc{Sherpa}~\cite{Sjostrand:2014zea, Bellm:2015jjp, Gleisberg:2008ta}, each with different shower ordering variables and evolution schemes. These differences lead to variations in the predicted jet structure and soft radiation profiles, which are constrained by data from LEP, HERA, and LHC measurements.

Thus, the parton shower evolution provides a crucial bridge between the hard, short-distance scattering and the long-distance, non-perturbative regime of QCD. It achieves this by resumming the dominant contributions from multiple soft and collinear gluon emissions, providing a realistic, exclusive final-state structure for collider observables.

\subsubsection{Hadronisation}
\label{subsubsection:hadronisation}

The hadronisation process bridges the gap between the perturbative domain of parton showers and the observable world of colour-neutral hadrons. Since QCD becomes strongly coupled at low energies, the transition from quarks and gluons to hadrons cannot be described reliably by perturbation theory. Instead, hadronisation must be modelled using phenomenological approaches that capture the essential features of confinement and the dynamics of colour neutralisation.

The most widely used hadronisation models are the Lund string model~\cite{Andersson:1983ia, Sjostrand:1984ic}, implemented in \textsc{Pythia}, and the cluster model~\cite{Webber:1983if, Bahr:2008tx}, used in \textsc{Herwig}. QCD inspires both models, but they rely on adjustable parameters that must be fitted to data from $e^+e^-$, $ep$, and hadron collider experiments.

In the Lund string model, the colour field between a quark and an antiquark is modelled as a one-dimensional relativistic string with a linear confinement potential. When a quark and an antiquark are pulled apart, the string stretches until it breaks by producing new $q\bar{q}$ pairs from the vacuum. This string-breaking process repeats, resulting in the formation of hadrons. The fragmentation function, which determines the energy and momentum distribution of the resulting hadrons, is parametrised and tuned to data. The string model naturally accounts for features such as longitudinal momentum correlations and the suppression of heavy-flavour production.

In contrast, the cluster model is based on preconfinement. At the end of the parton shower, colour-singlet clusters of partons are formed from colour-connected parton pairs. These clusters are then decayed isotropically into hadrons according to their mass and available phase space. This approach emphasises the colour structure emerging from the perturbative evolution and attempts to model hadronisation through a minimal set of assumptions. The cluster model tends to produce fewer long-range correlations compared to the string model.

Both models include dedicated treatments for baryon production, strangeness suppression, heavy-flavour hadronisation, and the modelling of soft radiation. Many of these aspects are controlled by non-perturbative parameters. These parameters are constrained by global fits to experimental data, often using measurements from LEP and SLD for $e^+e^-$ annihilation, HERA for $ep$ collisions, and LHC data for $pp$ processes~\cite{Buckley:2009bj, Skands:2010ak, Bellm:2019zci}.

Further developments include the incorporation of colour reconnection models, which allow for rearrangement of colour flows before hadronisation and can affect final-state observables such as jet shapes, underlying event structure, and particle multiplicities~\cite{Christiansen:2015yqa, Argyropoulos:2014zoa}. Although such effects are not derived from first principles, they are essential for achieving agreement with data.

Despite their phenomenological nature, hadronisation models are remarkably successful in describing a wide range of observables. Their robustness and tuning are critical for reliable Monte Carlo simulations of high-energy collisions. Ongoing efforts aim to constrain the models further using modern machine learning techniques and to enhance their theoretical underpinnings with insights from lattice QCD and effective field theories.

Following hadronisation, most of the produced hadrons are unstable and decay further into long-lived particles. Short-lived mesons and baryons, such as resonances and strange hadrons, undergo decays governed by the weak or strong interactions, depending on their quantum numbers and lifetimes. The final set of stable or quasi-stable particles typically includes charged pions (\( \pi^\pm \)), kaons (\( K^\pm \)), protons and antiprotons $( p, \bar{p})$, electrons $(e^\pm)$, muons $(\mu^\pm)$, photons $( \gamma)$, and neutrinos $(\nu_e, \nu_\mu, \nu_\tau)$.

These are the particles that ultimately reach the detector and form the basis for experimental measurements and event reconstruction. While the initial hard process may involve only a few partons (of order 2 to 4), the parton shower typically produces tens of radiated gluons and quarks. After hadronisation and decay, the multiplicity increases substantially. A typical LHC event at high energies results in hundreds to thousands of final-state particles, depending on the process and the phase-space region considered.

This proliferation of particles reflects the rich structure of QCD and the essential role of non-perturbative dynamics. Understanding the transition from few-particle perturbative events to high-multiplicity hadronic final states is central to the simulation and interpretation of collider data.

\subsubsection{Jets and Their Substructure}
\label{subsubsection:jets}

Jets are special among reconstructed objects at the LHC and deserve a separate discussion. They do not correspond to individual particles, but rather to collimated sprays of hadrons originating from the fragmentation of quarks and gluons. Furthermore, they are the most frequently observed, yet simultaneously the least well-understood, objects in collider physics. Their ubiquity and complexity demand a dedicated and detailed discussion.

Due to confinement in QCD, partons produced in the hard interaction and subsequent parton shower manifest as sprays of hadrons in the detector. The concept of a jet provides a means to reconstruct and interpret this collective behaviour using observables that approximate the properties of the original partons. Jets are reconstructed from final-state particles using clustering algorithms that combine energy deposits and tracks into a single object with a well-defined four-momentum.

\begin{figure}[ht]
	\centering
	\includegraphics[width=.7\textwidth]{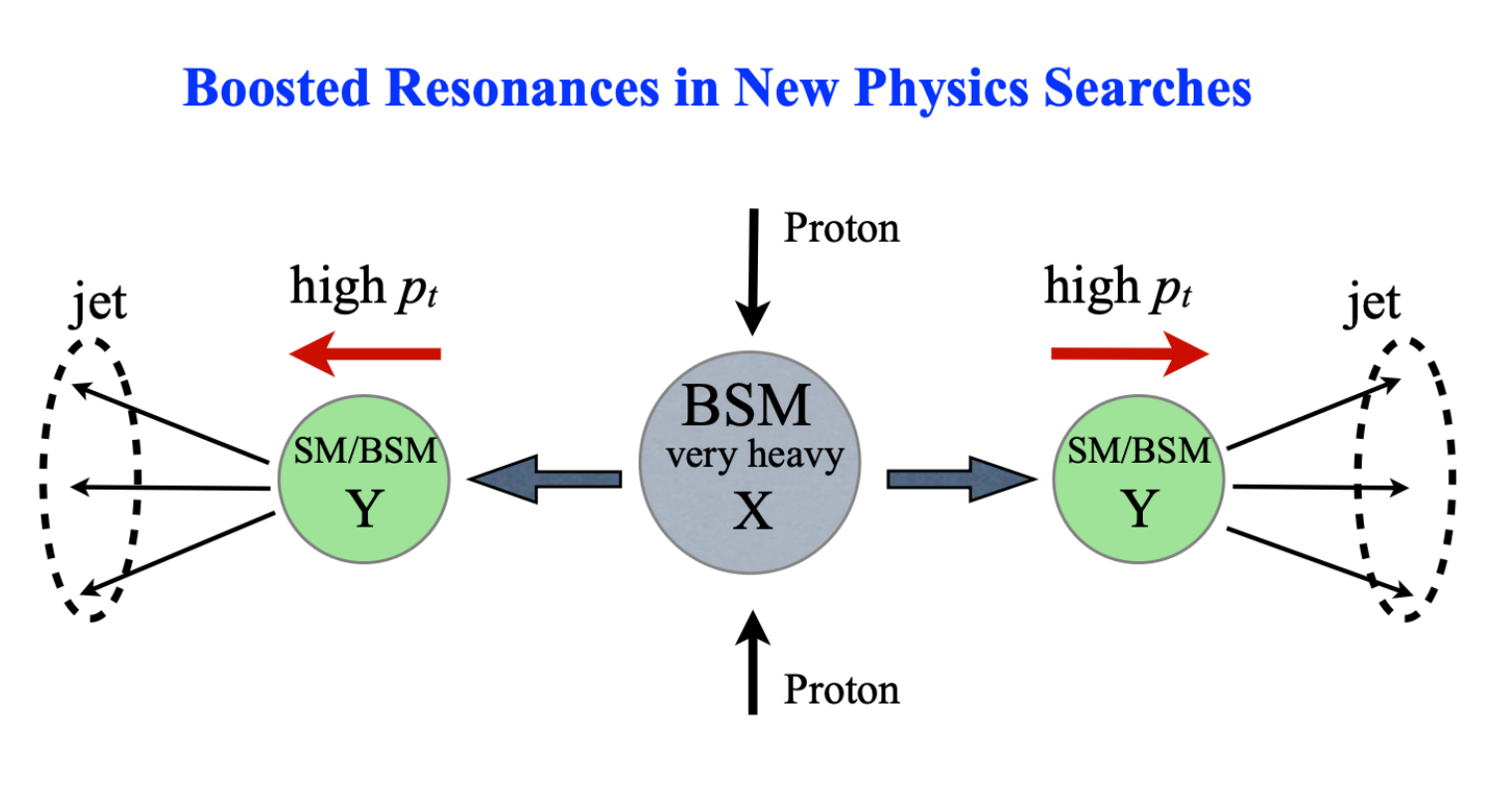}
	\caption{Generic interaction sequence for the search of a very BSM resonance that
decays into electroweak-scale particles that subsequently decay hadronically. The interaction produces a boosted resonance, which decays into two jets with highly collimated substructure. The kinematics of the jets are crucial for reconstructing the resonance mass and distinguishing it from background QCD jets.}
	\label{fig:boost_kinematics}
\end{figure}

Historically, fixed-cone algorithms were employed to group nearby particles within a cone of radius $ R $ in the $(\eta, \phi)$ plane. However, such algorithms were prone to infrared and collinear (IRC) unsafety, meaning that the addition of a soft particle or the collinear splitting of a hard particle could alter the jet structure in an unphysical way.

The development of IRC safe sequential recombination algorithms marked a major advance in jet physics. In these methods, each proto particle i is assigned a set of distances to every other proto particle j and the beam. The pairwise and beam distances are
\[
 d_{ij} = \min \left(p_{Ti}^{2p}, p_{Tj}^{2p} \right)~\frac{\Delta R_{ij}^2}{R^2}, \qquad d_{iB} = p_{Ti}^{2p},
\]
with $\Delta R_{ij}^2 = (\eta_i - \eta_j)^2 + (\phi_i - \phi_j)^2$ the rapidity azimuth separation and $R$ the jet radius parameter. A single exponent $p$ controls the ordering of the clustering and therefore the geometry of the resulting jets. When $(p=1)$, the measure reproduces the $k_T$ algorithm \cite{Catani:1993hr, Ellis:1993tq}, which merges the softest objects first and tends to form irregular jet boundaries that adapt to soft radiation. When $p=0$, one obtains the Cambridge–Aachen algorithm \cite{Dokshitzer:1997in}, which orders recombinations purely by angle and is well suited to substructure studies that exploit the angular hierarchy of QCD. When $p=-1$, the distance measure becomes anti-$k_T$ \cite{Cacciari:2008gp}, which effectively locks onto hard seeds so that soft particles are absorbed by the nearest hard core, producing nearly circular jets that are resilient to pile up and convenient for calibration.

The clustering proceeds by repeatedly identifying the smallest entry among all $d_{ij}$ and $d_{iB}$. If a pairwise distance is smallest, the corresponding objects are recombined, usually with the four-vector $E$ scheme addition, and the list of proto particles is updated. If the beam distance is smallest, the corresponding object is promoted to a final jet and removed from the list. The procedure is iterated until no proto particles remain. Because soft emissions carry small $p_T$ and collinear splittings preserve the measure, these algorithms are infrared and collinear safe by construction. The radius $R$ sets the effective catchment area of a jet. It should be chosen in relation to the transverse momentum scale of the signal and the expected level of soft contamination. In practice, anti-$k_T$ with a fixed $R$ has become the default at the LHC for inclusive jet reconstruction. At the same time, Cambridge-Aachen is often used as a basis for declustering and grooming in jet substructure analyses, and $k_T$ remains valuable when one wishes to expose the soft structure of an event.

At each iteration, the smallest of all distances $d_{ij}$ and $d_{iB}$ is selected. If a $d_{ij}$ is smallest, particles $i$ and $j$ are merged. If a $d_{iB}$ is smallest, particle $i$ is declared a final jet and removed from the list. This procedure is repeated until all particles are assigned to jets.

Among these, the anti-$k_T$ algorithm is currently the default choice at the LHC due to its resilience to pile-up and its production of nearly conical jets. These features simplify calibration and analysis in high-luminosity environments.

Jet substructure techniques were developed to improve the identification of hadronically decaying heavy particles, such as boosted top quarks and electroweak bosons, whose decay products become highly collimated and are often clustered into a single fat jet. The internal structure of such jets is then analysed to distinguish signal-like two- or three-prong topologies from background QCD jets. A central observable is N-subjettiness \cite{Thaler:2010tr, Thaler:2011gf}, which measures how well a jet can be resolved into N candidate axes. It is defined by
\[
\tau_N = \frac{1}{d_0} \sum_k p_{T,k} \, \min\{ \Delta R_{1,k}, \ldots, \Delta R_{N,k} \}, \qquad d_0 = \sum_k p_{T,k} ~ R,
\]
where the sum runs over constituents k, the $\Delta R_{a,k}$ denote distances to the candidate axes, and $R$ is the jet radius. Ratios such as $\tau_{21} = \tau_2/\tau_1$ efficiently separate two-prong decays from one-prong QCD jets. Energy correlation functions provide complementary information \cite{Larkoski:2013eya}, which use pairwise and higher-point correlations of constituent energies and angular separations to construct IRC-safe discriminants that are sensitive to the number of hard prongs. Another robust handle is the groomed jet mass, obtained by systematically removing soft contamination, ensuring that the resulting mass remains correlated with the parent particle mass \cite{Dasgupta:2013ihk}.

Mitigating soft radiation and pile-up is essential for these measurements; hence, grooming algorithms play a central role. Trimming removes subjets whose transverse momentum falls below a fixed fraction of the fat-jet \(p_T\) \cite{Krohn:2009th}. Pruning discards soft, large-angle branches encountered during reclustering, improving mass resolution in busy environments \cite{Ellis:2009su}. Soft drop declustering generalises these ideas by recursively undoing the clustering and imposing the condition
\[
\frac{\min(p_{T1}, p_{T2})}{p_{T1}+p_{T2}} > z_{\text{cut}} \left( \frac{\Delta R_{12}}{R} \right)^{\beta},
\]
thereby removing soft wide-angle radiation while retaining the hard two-prong core \cite{Larkoski:2014wba}. The choice of $z_{\text{cut}}$ and $\beta$ controls the aggressiveness and theoretical properties of the grooming.

Several dedicated taggers combine these ingredients into algorithms tailored to specific signals. The HEPTopTagger resolves three-prong substructure consistent with a hadronic top decay inside a large-radius jet. It searches for a hard, moderately symmetric three-prong configuration, applies mass-drop and filtering steps to stabilise the reconstructed top mass, and uses kinematic constraints to select top candidates \cite{Plehn:2010st, Plehn:2011sj}. Shower Deconstruction takes a complementary likelihood-based approach, assigning to each jet a weight that compares the probability for a signal-like shower history against a background-like one, computed from simplified analytic shower kernels and Sudakov factors \cite{Soper:2011cr, Soper:2012pb}. Both methods have demonstrated strong performance in boosted top and electroweak boson tagging, and they remain competitive benchmarks alongside observables such as N-subjettiness and energy correlation functions.

These techniques have significantly improved the identification of boosted hadronic decays of top quarks, $W$, $Z$, and Higgs bosons, and they are vital in searches for new heavy resonances. In particular, substructure information helps to isolate final states predicted in scenarios with heavy vector resonances, excited quarks, or composite Higgs sectors, where signal jets are typically more pronged and more massive than QCD backgrounds.

In recent years, machine learning approaches have transformed jet substructure studies. These include deep neural networks trained on jet images, graph neural networks acting on sets of constituents, and techniques such as particle flow networks~\cite{Komiske:2018cqr, Qu:2019gqs}. These approaches capture high-dimensional information and improve tagging performance beyond traditional observables.

On the theoretical side, the description of jet observables has been advanced by soft collinear effective theory, which provides a framework to factorise cross sections and resum logarithms associated with jet kinematics \cite{Becher:2014oda}, and by traditional perturbative QCD resummation based on factorisation and coherent branching. Canonical examples include the transverse momentum resummation of Collins, Soper and Sterman \cite{Collins:1984kg}, threshold resummation \cite{Sterman:1986aj, Catani:1989ne}, and all-order treatments of event shapes, jet vetoes, and non-global logarithms such as the CAESAR and BMS formalisms \cite{Banfi:2004yd, Dasgupta:2001sh, Banfi:2002hw, Banfi:2012jm}. These methods enable high precision predictions for jet masses, groomed observables, and jet veto cross sections, and they complement each other in the regions of phase space relevant to substructure measurements \cite{Bozzi:2005wk}.

Thus, jets and their substructure constitute a vital domain of collider physics, connecting the strong interaction dynamics of QCD with experimental observables. Their study continues to evolve through theoretical innovation, algorithmic development, and machine learning applications.

\section{Kinematic Observables and Event Selection}\label{sec3}

A central task in collider phenomenology is to define and measure kinematic observables that efficiently separate signal processes from Standard Model backgrounds. These observables are constructed from the four-momenta of reconstructed objects in the detector and exploit differences in production kinematics, resonance structure, and decay topology between competing hypotheses.

At a deeper level, these observables are engineered to expose the quantum numbers of the underlying interaction. Spin, parity, and gauge charges leave characteristic imprints in the event’s radiation pattern, so well-chosen kinematic variables turn those quantum numbers into measurable distributions.

Basic variables include the transverse momentum $p_T$ of an object, defined in Sec.~(\ref{sec1:subsec1}), the pseudorapidity $\eta$, and the azimuthal angle $\phi$. The invariant mass of a pair or system of objects,
\begin{equation}
  m_{ab} = \sqrt{(E_a+E_b)^2 - (\vec{p}_a + \vec{p}_b)^2},
\end{equation}
is a particularly powerful observable when searching for resonances such as the $Z$ boson, Higgs boson, or hypothetical heavy states like a $Z'$ \cite{ATLAS:2012yve, CMS:2012qbp, CMS:2016abv}. For final states with invisible particles, the transverse mass
\begin{equation}
  m_T = \sqrt{2 p_T^\ell E_T^{\mathrm{miss}} (1 - \cos \Delta\phi_{\ell,\,\mathrm{miss}})}
\end{equation}
provides discrimination between $W$ boson decays and backgrounds without genuine missing energy.

In multi-jet final states, the separation in rapidity–azimuth space,
\begin{equation}
  \Delta R_{ij} = \sqrt{(\eta_i - \eta_j)^2 + (\phi_i - \phi_j)^2},
\end{equation}
is widely used to define isolation cones around leptons and photons or to reconstruct jet substructure features \cite{Salam:2010nqg}. Large $\Delta R$ separations can signal boosted heavy particle decays, where decay products are collimated.

\paragraph{Azimuthal correlations in weak boson fusion Higgs production: a higher-level observable}

A canonical example of such a higher-level observable is the azimuthal separation of the two tagging jets in Higgs production via weak boson fusion (WBF),
\[
\Delta\phi_{jj} = |\phi_{j_1} - \phi_{j_2}| \in [0,\pi].
\]
In the Standard Model with a purely CP-even $HVV$ interaction, the $\Delta\phi_{jj}$ distribution is only mildly modulated and tends to be enhanced near $\pi$ once realistic selections are applied. If the Higgs couples through an admixture of CP-even and CP-odd operators, interference generates a characteristic phase and shape change. In the extreme CP-odd limit the distribution develops a near $\sin^2\!\Delta\phi_{jj}$ behaviour, suppressing events at $\Delta\phi_{jj}\simeq 0,\pi$ and enhancing the region around $\pi/2$ \cite{Plehn:2001nj, Hankele:2006ma, Klamke:2007cu, DelDuca:2006hk}. Because WBF features colour-singlet $t$-channel exchange and large rapidity separation between the tagging jets, the azimuthal correlation survives soft radiation and underlying event activity better than in gluon-fusion $H$+2-jet topologies \cite{DelDuca:2006hk}. This makes $\Delta\phi_{jj}$ in WBF a clean analyser of the CP structure of the $HVV$ vertex, directly connecting a measurable distribution to the quantum numbers of the resonance involved.

Experimentally, the extraction proceeds by selecting two high-$p_T$ jets with large rapidity separation and invariant mass, vetoing additional central jets to suppress QCD backgrounds, and reconstructing the Higgs decay (for example, $H\to\tau\tau$ or $H\to WW^*$). The $\Delta\phi_{jj}$ spectrum is then compared to templates corresponding to different CP hypotheses or effective operator admixtures, with systematic uncertainties constrained in control regions. ATLAS and CMS have already performed analyses along these lines, placing bounds on CP-odd admixtures in the $HVV$ interaction using WBF-enriched $H\to\tau\tau$ and $H\to WW^*$ final states \cite{ATLAS:2023dew, CMS:2021nnc}. Theoretical control of the observable is well developed, with NLO QCD corrections, parton-shower effects, and hadronisation found to preserve the discriminating power when standard WBF selections and central-jet vetoes are used \cite{Hankele:2006ma, Klamke:2007cu}.

\paragraph{Object tagging}
Object identification further enhances kinematic discrimination by turning radiation patterns and displaced decay topologies into robust classifiers of the initiating particle. Heavy-flavour tagging (b‑tagging) exploits long lifetimes of hadrons containing b quarks through impact‑parameter significance, secondary‑vertex reconstruction, and soft‑lepton information, and is a cornerstone of top and Higgs measurements as well as many BSM searches \cite{ATLAS:2015dex, CMS:2017wtu}. Closely related techniques enable charm tagging, with dedicated discriminants trained to separate c jets both from b jets and from light‑flavour jets, and tau‑tagging, which targets the narrow, low‑track‑multiplicity prongs and displaced secondary vertices characteristic of hadronic $\tau$ decays (see performance summaries in \cite{ATLAS:2015dex, CMS:2017wtu} and refs. therein).

Beyond flavour, modern taggers identify the internal multi‑prong structure of jets to tag boosted electroweak bosons and top quarks. Higgs tagging in the $H\to b\bar b$ channel pioneered the use of mass‑drop filtering and subjet $b$‑tagging to recover a resonant mass peak within a single large‑radius jet \cite{Butterworth:2008iy}. Hadronic $W/Z$ bosons and top quarks are tagged using groomed jet mass combined with prong‑sensitive observables such as $N$‑subjettiness ratios ($\tau_{21},~\tau_{32}$) or energy‑correlation functions, often within dedicated taggers like the HEPTopTagger \cite{Thaler:2010tr, Thaler:2011gf, Larkoski:2013eya, Plehn:2010st, Plehn:2011sj}. These methods provide powerful handles for measurements and searches that rely on boosted final states.

A complementary classification problem is quark–gluon tagging, which leverages the broader radiation pattern and higher particle multiplicity of gluon jets compared to quark jets. Discriminants built from charged‑particle multiplicity, jet width, particle $p_T$ spectra, and IRC‑safe shape observables have been developed and systematically studied \cite{Gallicchio:2011xq, Larkoski:2014gra}. While intrinsically less universal due to process and scale dependence, quark–gluon tagging is valuable for background suppression and for constraining theory systematics in data‑driven analyses.

Flavour tagging (b, c, $\tau$), boson and top tagging ($H$, $W$, $Z$, $t$), and quark–gluon discrimination convert detailed properties of jets and secondary vertices into physics labels that dramatically improve sensitivity. Combined with grooming and substructure observables \cite{Thaler:2010tr, Thaler:2011gf, Larkoski:2013eya, Larkoski:2014wba, Ellis:2009su, Krohn:2009th}, these taggers form the backbone of many collider searches and measurements, and their performance continues to advance through improved algorithms and high‑quality calibration on data (see e.g. the reviews \cite{Marzani:2019hun, Larkoski:2017jix}).

\vspace{0.3cm}
More sophisticated selection strategies use multivariate techniques such as boosted decision trees or neural networks \cite{Voss:2007jxm, Baldi:2014kfa}, trained on simulated signal and background samples to optimise separation in a high-dimensional space of observables. While these methods can offer substantial gains in sensitivity, their outputs are often still interpreted through the lens of the underlying kinematic variables to preserve physical insight.

Careful design of observables and selection criteria is essential for robust measurements and searches. They must balance statistical sensitivity, control over systematic uncertainties, and compatibility with theoretical predictions. The next section will describe how these selections feed into statistical inference frameworks used to quantify the presence or absence of a signal.

\section{Statistical Interpretation in Collider Phenomenology}\label{sec4}

Once observables and event selections have been defined, collider phenomenology proceeds to quantify how compatible the observed data are with different physics hypotheses. The standard language is that of likelihood functions $\mathcal{L}(\text{data}~|~\boldsymbol{\theta})$, which encode the probability of the data given model parameters $\boldsymbol{\theta}$. The parameter vector typically splits into a parameter of interest, such as a signal strength $\mu$ or a coupling, and a set of nuisance parameters $\boldsymbol{\nu}$ that represent systematic effects, including efficiencies, calibrations, and theoretical uncertainties.

\paragraph{Building the likelihood.}
Analyses typically construct a binned likelihood over one or more discriminant distributions. In a single channel with bins indexed by $i$, the expected yield in bin $i$ is $\lambda_i(\mu,\boldsymbol{\nu}) = \mu s_i(\boldsymbol{\nu}) + b_i(\boldsymbol{\nu})$, where $s_i$ and $b_i$ are the nominal signal and background expectations possibly modified by shape systematics. The observed counts $n_i$ are modelled with Poisson terms, $\prod_i \mathrm{Pois}(n_i~|~\lambda_i)$. Auxiliary measurements constrain nuisance parameters through Gaussian, log-normal, or gamma constraint terms that reflect prior knowledge from calibrations or control regions. Limited Monte Carlo statistics are often treated with the Barlow–Beeston prescription, which introduces per-bin nuisance parameters to propagate template statistical uncertainties \cite{Barlow:1993dm, Cranmer:2012sba}. Multi-channel fits generalise this construction to several signal and control regions with shared parameters and correlated systematics.

Templates that depend on $\boldsymbol{\nu}$ are implemented by morphing between up and down systematic variations, preserving normalisation and shape changes. Frameworks like HistFactory provide a declarative syntax for such models and handle correlations across regions and processes \cite{Cranmer:2012sba}. Unbinned likelihoods are also used when full event-by-event information is needed, for example, in precise mass measurements; however, binned templates are the workhorse for most searches.

\paragraph{Hypothesis testing.}
Discovery and exclusion questions are phrased as hypothesis tests. The background-only hypothesis $H_0$ corresponds to $\mu=0$, while the signal-plus-background hypothesis $H_1$ has $\mu>0$. A widely used test statistic is the profile likelihood ratio
\begin{equation}
q_\mu = -2\,\ln\frac{\mathcal{L}(\text{data}\,|\,\mu,\hat{\hat{\boldsymbol{\nu}}}_\mu)}{\mathcal{L}(\text{data}\,|\,\hat{\mu},\hat{\boldsymbol{\nu}})} ,
\end{equation}
where hats denote unconditional maximum likelihood estimators and double hats conditional estimators at fixed $\mu$ \cite{Cowan:2010js}. For discovery one tests $\mu=0$ and quotes a local significance $Z$ obtained from the $p$-value $p_0$ using the one-sided Gaussian mapping, $Z=\Phi^{-1}(1-p_0)$. Conventionally, $Z\simeq 5$ marks a discovery (often referred to in experimental publications as an ‘observation’).

When scanning across mass or other parameters, the \emph{look-elsewhere effect} dilutes the global significance. A common treatment is to correct the local $p$-value using trial factors estimated from ensembles or asymptotic approximations \cite{Gross:2010qma}. For exclusion, the LHC experiments employ the CL$_s$ construction \cite{Read:2002hq}, defined as $\mathrm{CL}_s = \mathrm{CL}_{s+b}/\mathrm{CL}_b$, which avoids excluding signals to which the analysis has poor sensitivity.

\paragraph{Intervals and parameter estimation.}
Confidence intervals for $\mu$ and other physics parameters follow from likelihood scans. In the asymptotic regime, Wilks’ theorem implies that $-2\Delta\ln\mathcal{L}$ is $\chi^2$-distributed, so that one- and two-parameter intervals correspond to fixed contours of the profile likelihood. The \emph{Asimov} dataset provides median expected sensitivities without running ensembles \cite{Cowan:2010js}. In small samples or strongly non-Gaussian problems, frequentist intervals are constructed with ensembles or by adopting the Feldman–Cousins ordering to ensure proper coverage \cite{Feldman:1997qc}. Bayesian credible intervals are also used in some contexts by marginalising over $\boldsymbol{\nu}$ with specified priors; the numerical work is performed with Markov chain or nested sampling methods.

\paragraph{Systematic uncertainties and control regions.}
Nuisance parameters represent experimental and theoretical systematics and are constrained by auxiliary terms and by data in control regions. Typical experimental uncertainties include luminosity, trigger and reconstruction efficiencies, energy scales and resolutions, and flavour tagging calibrations. Theory uncertainties cover cross sections, parton shower and hadronisation modelling, and PDF and scale variations. Shape systematics are introduced through alternative templates and correlated across processes and regions following the underlying physics. Backgrounds are often estimated with \emph{transfer factors} from control to signal regions or with sideband fits. Validation regions, kinematically close to the signal region but signal-depleted, test the modelling before unblinding.

Correlations matter. A single nuisance, for example, the jet energy scale, may affect many bins and regions simultaneously. Over-constraining or double-counting must be avoided. Post-fit nuisance \emph{pulls} and \emph{impacts} provide diagnostics of how the data constrain systematics and which uncertainties dominate the result.

\paragraph{Goodness of fit and model comparison.}
Beyond discovery and exclusion, one should check whether the fitted model describes the data. Goodness-of-fit tests use $\chi^2$-like statistics on binned residuals, saturated likelihood ratios, or tests based on probability integral transforms. Model comparison between non-nested hypotheses can be addressed with information criteria or dedicated likelihood ratio tests constructed on common parameterisations.

\paragraph{Tools and workflows.}
The statistical machinery is implemented in public tools used by experiments and the community. RooFit and RooStats provide likelihood modelling, test statistics, and interval construction \cite{Verkerke:2003ir, Moneta:2010pm}. HistFactory defines templated analyses with constraints and correlations \cite{Cranmer:2012sba}. The CMS \textsc{Combine} framework integrates these ingredients in a reproducible workflow \cite{CMS-NOTE-2011-005}. Increasingly, results are published with simplified likelihoods and covariance information on HEPData \cite{Maguire:2017ypu} to enable reinterpretation by external users.

\vspace{0.3cm}
Statistical interpretation connects measured event distributions to physics conclusions. Likelihood-based tests quantify discovery or exclusion, intervals constrain parameters, and a disciplined treatment of nuisance parameters propagates experimental and theoretical uncertainties. Together, these elements ensure that collider measurements and searches yield results that are comparable across analyses and reusable in global interpretations.

\section{Examples of Searches for electroweak- to TeV-scale New Physics}\label{sec5}

In this section, we illustrate how the elements introduced above come together in representative collider analyses. Each example identifies a set of observables, defines an event selection, and applies the statistical machinery of Sec.~\ref{sec4} to reach physics conclusions. The goal is to show the complete flow from raw objects to a quantitative statement, discovery or limit, and to emphasise the connection between observables and the underlying quantum numbers.

\paragraph{Higgs boson discovery in $H\to\gamma\gamma$}
The diphoton channel provided one of the cleanest discovery modes for the Higgs boson at the LHC \cite{ATLAS:2012yve, CMS:2012qbp}. The key observable is the diphoton invariant mass, $m_{\gamma\gamma}$, reconstructed from two isolated, high-quality photons. The signal appears as a narrow peak on top of a smooth, falling continuum from prompt QCD diphoton production and from $\gamma$+jet and dijet events where jets fake photons. Event selection requires two isolated photons with transverse energies above thresholds tied to $m_{\gamma\gamma}$, and tight identification based on shower shapes and track isolation. Categories based on photon kinematics and event topology, including a vector boson fusion-enriched class with two forward jets, increase sensitivity by exploiting differences in mass resolution and signal-to-background ratios.

The background shape in $m_{\gamma\gamma}$ is modelled by analytic functions fitted to the sidebands. The signal shape is taken from a simulation and calibrated with control samples. A binned likelihood in $m_{\gamma\gamma}$, with nuisance parameters for energy scale, resolution, and background functional form, is used to test the background-only hypothesis and to extract the signal strength. The discovery was established by combining categories and datasets, resulting in a local significance exceeding five standard deviations near $m_{\gamma\gamma}\simeq 125\,\text{GeV}$. Differential measurements of $p_T^H$ and event categories sensitive to production modes have since enabled precision tests of the Higgs couplings and CP properties.

\paragraph{$Z'\to \ell^+\ell^-$ bump hunting}
Heavy neutral gauge bosons are benchmark signatures of many extensions of the Standard Model. The cleanest search utilises high-mass opposite-sign dileptons, $e^+e^-$ and $\mu^+\mu^-$, as well as other processes. The central observable is the dilepton invariant mass, $m_{\ell\ell}$, reconstructed with excellent resolution, up to the multi-TeV range. Event selection requires two isolated, high $p_T$ leptons, with quality and impact parameter requirements, as well as fiducial cuts in $\eta$. The dominant background is Drell–Yan production, with smaller contributions from $t\bar t$, dibosons, and instrumental backgrounds.

The analysis is a textbook example of a bump hunt. One compares the observed $m_{\ell\ell}$ spectrum with the smoothly falling expectation and tests narrow resonance hypotheses at each mass point. The test statistic is constructed from the profile likelihood in binned $m_{\ell\ell}$, with systematic uncertainties on efficiencies, energy or momentum scales, and background modelling profiled in the fit. The look elsewhere effect is accounted for when quoting global significances in wide scans \cite{Gross:2010qma}. In the absence of a significant excess, one sets limits on the cross-section times branching ratio as a function of mass, which are then mapped to model parameters for specific $Z'$ benchmarks. Recent LHC searches exclude sequential Standard Model-like $Z'$ bosons with masses above the 5 TeV scale \cite{CMS:2016abv}.

\paragraph{Constraining new physics with effective field theory in multi-boson production}
When new physics is heavy compared to the LHC energy, its leading effects can be captured by higher-dimensional operators in the Standard Model Effective Field Theory. Collider-sensitive operators modify the kinematics and rates of diboson and vector boson scattering processes, for example, $pp\to W^+W^-,~ WZ,~ ZZ$ and electroweak $VVjj$ production. The most sensitive observables are the tails of transverse momentum and invariant mass distributions, such as $p_T(V)$, $m_{VV}$, and in vector boson fusion topologies the dijet invariant mass $m_{jj}$, the rapidity separation $\Delta y_{jj}$, and azimuthal correlations like $\Delta\phi_{jj}$. These choices are driven by the energy growth of specific SMEFT amplitudes and by the clean colour singlet radiation pattern in vector boson fusion and scattering.

A representative analysis proceeds by defining signal regions enriched in the target topology, for instance, semileptonic $WZ\to \ell\nu jj$ or fully leptonic $ZZ\to 4\ell$, with control regions constraining $t\bar t$ and $W$+jets backgrounds. The likelihood combines several binned distributions, often the leading boson $p_T$ or $m_{VV}$, and is built as in Sec.~\ref{sec4} with nuisance parameters for experimental and theory systematics. Limits are set on SMEFT Wilson coefficients, either one at a time or in small sets, using the CL$_s$ method. Care is taken to remain in the regime where the EFT expansion is valid, which can be enforced by restricting to bins below an energy cutoff or by checking that the inferred coefficients yield perturbative predictions across the fit range. Dedicated measurements of vector boson scattering with two forward tagging jets and a central jet veto are particularly powerful for anomalous quartic gauge couplings.

On the theory side, one typically adopts the Warsaw basis of dimension six operators \cite{Grzadkowski:2010es} or equivalent parameterisations, and matches collider observables to Wilson coefficients using tools such as \textsc{FeynRules} and \textsc{MadGraph5\_aMC@NLO}. Global reviews provide guidance on operator choices and fit strategies \cite{Brivio:2019myy}. The LHC experiments have reported competitive bounds on triple and quartic gauge couplings, as well as on Higgs electroweak interactions, using these methods, with steadily improving precision as the datasets grow.

\paragraph{Remarks on reinterpretation and combinations.}
The three case studies above also illustrate the importance of publishing results in a form that enables reinterpretation. Shape-based limits and best-fit signal strengths are most useful when accompanied by simplified likelihoods and covariance information on HEPData \cite{Maguire:2017ypu}. This enables external users to test new models against existing measurements, combine channels, and consistently propagate correlated uncertainties.

\section{Conclusions}
\label{sec:conclusions}

Collider phenomenology provides the essential bridge between theory and experiment, translating the fundamental Lagrangian of the Standard Model and its possible extensions into measurable predictions that can be tested with collider data. We have attempted to outline the main components of this process: the definition of observables sensitive to the quantum numbers of the particles and interactions under study, the design of event selections to isolate the relevant processes from often overwhelming backgrounds, the use of advanced object reconstruction and tagging techniques, and the statistical methods that connect observed distributions to physics conclusions. We have illustrated these steps with representative case studies, ranging from the discovery of the Higgs boson, to high-mass resonance searches, to the extraction of bounds on new interactions in the framework of effective field theory.

A recurring theme has been the close interplay between theory and experiment. The design of observables is guided by quantum field theory expectations for the production and decay of particles, as well as by our understanding of QCD radiation patterns and detector effects. Event selections and tagging algorithms exploit these features to enhance signal-to-background ratios, while statistical inference tools translate the measured spectra into quantitative statements about model parameters or the presence of new phenomena. The examples discussed also illustrate how collider phenomenology adapts to different energy scales and luminosities, from precision studies at the electroweak scale to searches for new physics at the multi-TeV frontier.

As the LHC continues to collect data at higher energies and luminosities, and as preparations intensify for future colliders, the methods of collider phenomenology will face both challenges and opportunities. On the one hand, larger datasets enable more differential and multidimensional analyses, increasing sensitivity to subtle effects in both Standard Model and beyond-the-Standard-Model physics. On the other hand, the control of systematic uncertainties, experimental, theoretical, and statistical, becomes ever more critical, requiring continual refinement of analysis techniques, Monte Carlo modelling, and detector calibration.

Looking ahead, the integration of collider phenomenology with global analyses, incorporating results from flavour physics, astroparticle observations, and cosmology, will be key to building a coherent picture of fundamental interactions. The availability of open data products, such as simplified likelihoods on HEPData, will further enhance the reinterpretability and impact of collider measurements, allowing the community to test a wide array of new theories without repeating the complete experimental analysis.

Past and present colliders have been extraordinarily successful, from the early electron–positron and proton–antiproton machines to the Tevatron and the LHC, each step refining our understanding of the constituents and forces of nature. A next generation of colliders, whether precision facilities that harvest enormous samples at the electroweak scale or energy‑frontier machines that push well beyond the TeV range, would continue this trajectory. Higher precision will stress the Standard Model in new ways, reveal minor deviations in couplings and rare processes, and overconstrain the theory. Higher energies open qualitatively new final states and extend direct sensitivity to heavy particles and new interactions. To fully exploit either path, we must keep improving both our theoretical tools and our experimental methods, from higher-order and resummed calculations, global PDF determinations, parton showers and hadronisation models, and robust uncertainty quantification, to advances in detectors, triggering, pileup mitigation, reconstruction, tagging, and analysis strategies, including modern machine learning. With these developments, a future collider programme can deliver decisive advances in our understanding of the Higgs sector, the origin of mass scales, and the possible structure of physics beyond the Standard Model.

Collider phenomenology is a mature yet constantly evolving discipline. It combines theoretical insight, experimental ingenuity, and statistical rigour to maximise the physics output of collider experiments. By continually refining its tools and methods and by embracing open and collaborative practices, it will remain central to the quest for understanding the fundamental constituents of nature and their interactions.

%%%%%%%%%%%%%%%%%%%%%%%%%%%%%%%%%%%%%%%%%
%% Mandatory: Bibliography using bibtex 
\bibliographystyle{Numbered-Style} %% for Numbered Reference Style
\bibliography{reference}

\end{document}